\documentclass[prb,aps,twocolumn,superscriptaddress,showpacs]{revtex4}
\usepackage[dvips]{graphicx}
\usepackage{amsmath}
\usepackage{epsfig}
\usepackage{color}

\newcommand{\BFCA}{Ba(Fe$_{1-x}$Co$_x$)$_2$As$_2$ }
\newcommand{\kk}{{\bf k}}
\begin{document}

\title{The effect of disorder on electronic Raman scattering in the superconducting state of iron pnictides}

\author{G. R. Boyd}
\author{P. J. Hirschfeld}
\affiliation{Department of Physics, University of Florida,
    Gainesville, FL 32611, USA}
\author{T.P.Devereaux}
\affiliation{Stanford Institute for Materials and Energy Science, SLAC National Accelerator Laboratory, Menlo Park, CA 94025, USA and Geballe Lab for Advanced Materials, Stanford University, Stanford, CA 94305, USA}

\begin{abstract}
 Electronic Raman scattering measures a polarization-dependent scattering intensity which can provide information about the location of nodes in the energy gap of an unconventional superconductor as well as its overall symmetry. In this paper, we calculate the Raman intensity in the presence of disorder for several models of the iron pnictide superconducting state. We include, for completeness, $d$-wave and isotropic $s_\pm$ responses in addition to more realistic extended $s_\pm$ superconducting gaps.  The effect of disorder is modeled using a self-consistent $T$-matrix approximation, and is studied in the limits of isotropic and intraband-only scattering.  We show how recent experiments on \BFCA  may be consistent with ``node lifting" by intraband disorder.
\end{abstract}

\maketitle

\section{Introduction}

Electronic Raman scattering is sensitive to low-lying excitations and
can be performed for various polarizations of the incoming and outgoing photons. This polarization dependence preferentially samples different parts of the Brillouin zone, making Raman scattering  an important tool with which to clarify the location of nodes and the symmetry of the gap in the superconducting state\cite{DevereauxRMP}. Muschler et al.\cite{Muschler} have recently performed Raman scattering measurements on single crystals of \BFCA (122) for two different concentrations of Co. The data demonstrate the existence of low energy quasiparticle excitations in the superconducting state. These quasiparticles could be a result of pairbreaking effects or due to the presence of  nodes in the superconducting gap.
 In this paper we study the polarization dependent electronic Raman response in the presence of disorder in order to distinguish between these two possibilities.

Thus far, other experimental probes have not presented a convincing determination of a universal gap structure in
the Fe-pnictides \cite{DCJohnston,PaglioneGreene}. This may be due to variations in sample quality and resolutions issues of different measurement
techniques, but there is increasing speculation that these systems may possess a strong sensitivity of both the electronic structure
and the  pair state to small changes which affect electronic structure\cite{kemper10}.  This is because the Fermi surface consists of several
nearly compensated electron and hole pockets, and because the pairing state is probably of an extended-$s$ type
 which may possess ``accidental" nodes or deep minima, i.e. structures depending on the details of the pairing
 interaction rather than the symmetry class.
Nuclear magnetic resonance  (NMR) studies
\cite{ref:RKlingeler,ref:Grafe,ref:Ahilan,ref:Nakai}  showed a
$T^3$ spin lattice relaxation rate reminiscent of a  gap with
nodes.  ARPES measurements on single crystals of 122-type
materials
\cite{ref:Zhao,ref:Ding,ref:Kondo,ref:Evtushinsky,ref:Nakayama,ref:Hasan}
measured the gap reporting isotropic or nearly isotropic gaps on
all Fermi surface sheets. Penetration depth measurements \cite{ref:Hashimoto,ref:Malone,ref:Martin,ref:Hashimoto2,ref:Gordon,ref:Gordon2,ref:Fletcher}
have been fit both to an activated $T$-dependence,
indicative of a fully gapped state, and low-$T$ power laws, indicative of nodes in the superconducting gap.
It is possible that these differences reflect genuinely  different
ground states in different materials due to intrinsic differences in the pairing, but it is important to disentangle these
effects from those which arise from disorder, and in particular to distinguish between a disordered fully gapped state and a nodal state.

A popular candidate for the ground state of ferropnictides is the so-called $s_\pm$ state proposed by Mazin et al. \cite{ref:Mazin_Spm}.  In this state, found within a simple spin fluctuation model with strong interband scattering
 between nearly nested electron and hole pockets, the gap is  isotropic on both electron and hole pockets  but with a sign change between the two.  Further theoretical work\cite{ref:Wang_nodal_gapped,chubukov,r_thomale_09,s_graser_09,k_kuroki_09,k_kuroki_08,kemper10} considered spin fluctuation pairing using band structures derived from tight binding fits to density functional theory results, and found  $s_\pm$-type states
  with highly anisotropic gaps, particularly on the electron sheets. We refer to this general class of states, which are
  in the same symmetry class as ordinary $s$-wave and isotropic $s_\pm$ states, as ``extended-$s$" states here.  These calculations also reported   the near degeneracy of  $d$-wave gaps in certain situations. The effect of nonmagnetic disorder on an $s_\pm$, $d$-wave, and nodal extended-$s$ is different and thus may help narrow the field of candidate superconducting gaps. In particular, it has been shown that for the accidental nodes of an extended $s$-wave state, intraband disorder can 'lift' the nodes, resulting in a fully-gapped state\cite{ref:vivek}.  On the other hand, if disorder is of the interband type, low-energy impurity midgap states can be created \cite{ref:Yu, ref:Shiba, ref:Hussey,GolubovMazin,SengaKontani} similar to Yu-Shiba states due to magnetic impurities in ordinary superconductors.

One complication is that changing the doping does not have the simple effect of creating more point-like disorder. To what extent a dopant is charged or magnetic, long or short ranged, and what changes it makes in band structure upon doping requires careful study\cite{AFKemperCo,SawatzkyCo} and is difficult to include consistently in a theory capable of calculating observable quantities. In the most naive approach, we will consider doping as loosely related to the concentration of scatterers in the sense that higher doping is a dirtier sample, with the caveat that  effects on the electronic structure and pair interaction may also be present. Here we report the results of a calculation of the Raman intensity for $s_\pm$, $d$-wave, and experimentally inspired extended s-wave superconducting gaps, including disorder using a self-consistent T-matrix approximation (SCTMA). We show that for unitary scatterers, intra-band scattering will average the gap and 'lift' the nodes of an extended s-wave state, leaving a fully developed gap for higher scattering rates due to disorder. When a strong interband scattering component is present, the creation of an impurity band competes with the tendency to lift the accidental gap nodes. In contrast to the extended-s scenario, in an isotropic $s_\pm$ model, disorder causes the creation of low energy quasiparticles mimicking power law behavior of nodal states for some probes. One important difference between such a sign-changing isotropic state and a nodal extended-s state, then, is that the addition of disorder fills in the gap as opposed to possibly creating it.

The form of the Raman spectrum will be influenced additionally by inelastic scattering processes.  We expect these
to be largely frozen out in the superconducting state at low energies, but they will be important for a proper
treatment of the normal state spectrum and for energies near the maximum gap.  Effects of this type will be treated elsewhere.\cite{Kempertobepublished}

Our paper is organized as follows: in the first section we describe the theoretical background necessary to undertake this study. In the next section, we describe simple one and two band results for the $d$-wave and $s_\pm$ cases to help ground our understanding. Finally, we show the results from studying anisotropic $s$-wave gaps on all four Fermi sheets which we suggest captures all the essential qualitative features of the real Co-doped 122s.

\section{Theory of Electronic Raman Scattering with disorder}
\label{theory}

Raman scattering is the inelastic scattering of polarized light from a material  (for a review see Ref.\onlinecite{DevereauxRMP}). The cross section of the scattered light is proportional to\cite{DevKampf}
\[
S_{\gamma,\gamma}= \frac{\omega_{sctr}}{\omega_{inc}}\frac{e^2}{mc^2}[1+n_B(\omega)]\frac{1}{\pi}Im \chi_{\gamma,\gamma}
\]
where
\begin{equation}
\chi_{\gamma,\gamma}(\omega)=\int_0^{\beta}d\tau e^{-i\omega_m\tau}
\langle T_{\tau}\tilde\rho_{\gamma}(\tau)\tilde\rho_{\gamma}(0)\rangle
\mid_{i\omega_m\rightarrow w+i\delta},
\end{equation}
and
\[
\chi(q \rightarrow 0 ) = T \sum_n \sum_\kk Tr (\gamma_\kk^2 \tau_3 G(\kk,i\omega_n) \tau_3 G(\kk,i \omega_n + i \Omega_m))
\]

Here $\chi_{\gamma,\gamma}$ is the Raman effective density-density correlation function for symmetry channel $\gamma$. The vertex, $\gamma_{\kk}$, accounts for the interaction of polarized light with charge density.  The expression for the Raman effective density is $\rho=\sum_\kk \gamma_\kk c^{\dag}_{\kk,\sigma} c_{\kk,\sigma}$.
The full matrix Green's function in the
presence of scattering in the superconducting state is
\[
G(k, \omega) =\frac{\tilde{\omega}\tau_0
+\tilde{\epsilon_k}\tau_3+\tilde{\Delta_{\kk}} \tau_1 }{
\tilde{\omega}^2-\tilde{\epsilon_\kk}^2-\tilde{\Delta_\kk}^2},
\]
where $\tilde{\omega}\equiv \omega-\Sigma_0$,
$\tilde{\epsilon_\kk}\equiv \epsilon_\kk+\Sigma_3$, $\tilde{\Delta_\kk}
\equiv \Delta_\kk+\Sigma_1$, and the $\Sigma_\alpha$ are the
components of the disorder self-energy proportional to the Pauli matrices
$\tau_\alpha$ in particle-hole (Nambu) space.

Generally, the vertex is determined by both density and current matrix elements
between the conduction band and the excited states.
However, in situations where one is interested in qualitative results like the present one,
model Raman vertices classified by symmetry can be employed.  For a crystal with $D_{4h}$ tetragonal symmetry, in-plane charge fluctuations transform according to the full symmetry of the lattice (irreducible representation $A_{1g}$), and do not
change sign upon 90 degree rotation, or lower symmetry, as opposed to the
$B_{1g}$ and $B_{2g}$ symmetry classes which change sign. Raman scattering probes long
wavelength charge fluctuations. The $B_{1g}$ and $B_{2g}$ charge
densities must average to zero within each unit cell, and are not
 coupled via the long-range Coulomb interaction. Thus the  $B_{1g}$ and $B_{2g}$ Raman responses for a multi-band
system consist of the sum of the contributions from each band.
However, $A_{1g}$ fluctuations need not vanish over the unit cell,
and therefore they can couple to isotropic charge density, giving rise
the finite backflow.  Therefore vertex corrections of both Coulomb and impurity type must be included in the $A_{1g}$ channel, which therefore becomes extremely difficult to treat consistently with disorder.  We do not treat this polarization channel in this work. Finally, we neglect direct interband and resonant contributions to the Raman vertices. In this approximation, the vertex for Raman scattering measures the effective mass around the FS. We use the notation that $\vec{e}$ is a polarization vector.
\[
\gamma_\kk = m \sum_{ab} {e_a} \frac{\partial^2 \epsilon}{\partial k_a\partial k_b}{e_b}
\]

In a square lattice the $B_{1g}$ and $B_{2g}$ polarizations, the vertices can be expanded as:
\[
B_{1g}: ~\gamma_{\bf k}=  \cos (k_x) - \cos(k_y)
\]
and
\[
B_{2g}: ~ \gamma_{\bf k}=\sin(k_x)\sin(k_y)
\]
which along a circular Fermi surface become $\cos(2\theta)$ and $\sin(2\theta)$, respectively. It should be noted here that in the iron-pnictides use of the 1-Fe vs. the 2-Fe Brillouin zone will interchange $B_{1g}$ and $B_{2g}$ symmetries.

\begin{figure}[h]
\includegraphics[width=80mm]{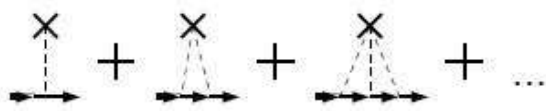}
\caption{ Diagrams representing the self-consistent T-matrix approximation (SCTMA). The dashed lines are scattering off an impurity, and the single line is the self-consistent green's function.}
\label{Tmat}
\end{figure}

To include the effect of disorder we self-consistently solve for the self energies by including all scatterings off a single impurity and performing a disorder average. This is represented diagrammatically in Fig.  \ref{Tmat}.  The
$T$-matrix can be defined as
\begin{equation}
G = G_0 + G_0 T G_0 + \cdots
\end{equation}

Using the Nambu notation, the single-particle self energy in a
superconductor can be decomposed as:
\begin{equation}
\tilde \Sigma(\kk,\omega)= \sum_\alpha \Sigma_\alpha(\kk,\omega) \tilde \tau_\alpha \,,
\end{equation}
where $\tilde \tau_\alpha$ are the Pauli matrices and $\tilde \tau_0$ is the unit matrix.
Note that the band index is implicitly contained in the $\kk$ index since we restrict pairing
to individual Fermi surface sheets.
Treating impurity-scattering in T-matrix approximation gives rise to
the following self-energy
\begin{equation}
\tilde \Sigma(\kk,\omega)= n_i \tilde T_{\kk \kk}(\omega),
\end{equation}
where $n_i$ is the impurity concentration and $T_{kk}(\omega)$ is the
diagonal element of the T-matrix
\begin{equation}
\label{eq:TMatrixk}
\tilde T_{\kk \kk'}(\omega)= V_{\kk \kk'} \tilde \tau_3
+ \sum_{\kk''} V_{\kk \kk''} \tilde \tau_3 \tilde G(\kk'',\omega) \tilde T_{\kk'' \kk'}(\omega)
\,.
\end{equation}
We define
\[
\Gamma=\frac{n_i n}{\pi N_0}
\]
where $n_i$ is the density of impurities,  \textit{n} of electrons, and $N_0$ the density of states at the fermi level.
For a constant potential, the case which we will consider in this paper, this expression becomes the series:
\[
\tilde \Sigma(\omega)= n_i V_0 \tilde \tau_3 \sum_n (\sum_{\kk'}   G(\kk',\omega) V_0 \tilde  \tau_3)^n
\]
Later we will restrict $V_{\kk \kk''}$ to be constant for particular sets of momenta, either to allow transitions between all the Fermi sheets or to restrict transitions to remain within Fermi sheets.

The self-energy $\tilde \Sigma(k,\omega)$ has to be solved
self-consistently in combination with the single-particle Green's function
\begin{equation}
\tilde G(\kk,\omega)^{-1}=\tilde G_0(\kk,\omega)^{-1}
-\tilde \Sigma(\omega).
\end{equation}

After solving for the self-energies, we insert them into the general expression for the Raman response.
Beginning with a spectral representation:
\[
G(\kk,i \omega_n)=\int dx \left(\frac{-1}{\pi}\right) \frac{Im G(\kk,x)}{i \omega_n - x}
\]
we arrive at a zero-temperature long-wavelength form,
\[
Im \chi_{\gamma,\gamma}(\Omega)=\sum_\kk \int^0_{-\Omega} dx \frac{1}{\pi} \gamma_\kk^2 Tr [ImG(\kk,x) \tau_3 Im G(\kk,x+\Omega) \tau_3 ]. \]
In terms of retarded and advanced Green's functions, $Im G = \frac{G^R-G^A}{2 i}$:
\[
Im \chi_{\gamma,\gamma}(\Omega)= \frac{1}{4 \pi} \langle N(\phi) Im \int d \xi \gamma^2_{\phi} \int_0^{\Omega} dx
\]\[
[F^{RR}(x-\Omega,x)-F^{RA}(x-\Omega,x)\]\[
-F^{AR}(x-\Omega,x)+F^{AA}(x-\Omega,x)] \rangle_{\phi}
\]
where
\[
F^{a,b}=Tr[G^{a}(\kk,x-\Omega) \tau_3 G^{b}(\kk,x) \tau_3 ] \; \;a,b=A,R
\]
We have taken $N(\phi)=N_0$ in this paper. The angular brackets denote an average over the angle $\phi$ around the Fermi surfaces.

We note here that for crossed polarizations selected $B_{1g}$ and $B_{2g}$ channels, the Raman response is additive for a many-sheeted Fermi surface. We also note that in the present approach we have neglected $T$-matrix impurity vertex corrections for the Raman response. While for general momentum-dependent $T_{\bf k,k^{\prime}}$, impurity vertex corrections are necessary for all channels, for momentum-independent scattering vertex corrections do not essentially modify the response for $B_{1g}$ and $B_{2g}$ polarizations\cite{DevKampf}.

\section{Raman Scattering for dirty $d$-wave and $s_\pm$-wave cases}
\label{dwaveswave}

First we present results for $d$-wave and $s_\pm$-wave superconducting gaps. Early on, the $s_\pm$ was proposed as a candidate for the gap structure in the iron-pnictides\cite{ref:Mazin_Spm}. Indeed, for certain materials parameters, in particular the case where a Fermi surface pocket at $(\pi,\pi)$ appears, this state is crudely consistent with multiorbital spin-fluctuation calculations\cite{ref:Wang_nodal_gapped,chubukov,r_thomale_09,s_graser_09,k_kuroki_09,k_kuroki_08,kemper10}.  The qualitative features of the  Raman response for such a state can be understood from a minimal two-Fermi-sheet model.

 NMR and penetration depth measurements in some Fe-pnictide materials displayed power-law temperature dependences suggestive of nodes in a single band model\cite{ref:RKlingeler,ref:Grafe,ref:Ahilan,ref:Nakai,ref:Hashimoto,ref:Malone,ref:Martin,ref:Hashimoto2,ref:Gordon,ref:Gordon2,ref:Fletcher}. The phenomenology was similar to what was seen in cuprates, so it is useful to include a study of the $d$-wave case for comparison, to how the presence of a node in the superconducting gap manifests itself in the Raman response. The $d$-wave Raman response has been previously studied, for example in Refs. \onlinecite{DevEinzel} \onlinecite{DevKampf}, and \onlinecite{TomPRL74}.

\begin{figure}[h]
\begin{flushleft}
\includegraphics[width=0.75\columnwidth,angle=270]{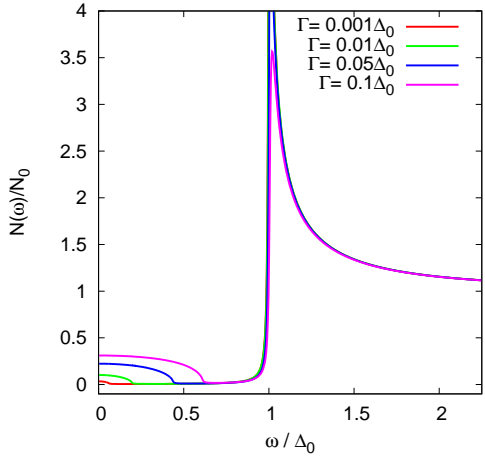}
\caption{(Color online) Quasiparticle density of states $N(\omega)$ for  isotropic $s_\pm$ state, normalized to normal state density of states $N_0$ vs. $\omega/\Delta_0$, where $\pm\Delta_0$ is the value of the gap on  hole and electron sheets.   $N_0$  is assumed constant on
all Fermi sheets.  Shown are  various interband impurity scattering rates $\Gamma$ in units of $\Delta_0$. }
\label{DOSSpm}
\end{flushleft}
\end{figure}

\begin{figure}[h]
\includegraphics[width=0.75\columnwidth,angle=270]{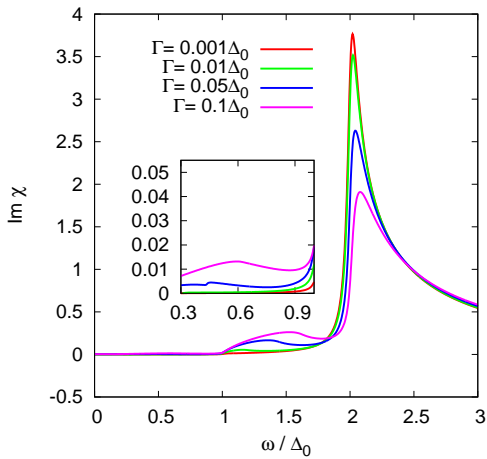}
\caption{(Color online) Raman response Im $\chi(\omega)$ vs. $\omega/\Delta$ for both $B_{1g}$ and $B_{2g}$ polarizations for an $s_\pm$ state as in Fig. \ref{DOSSpm} for various interband impurity scattering rates $\Gamma$, in units of $\Delta$ . }
\label{RamanSpm}
\end{figure}

We first consider the $s_\pm$ state on two circular Fermi sheets, each with an isotropic constant gap in the superconducting state $\Delta_{\pm}=\pm\Delta_0$. The gaps  differ by a minus sign, so inter-band scattering gives rise to pair-breaking and violates Anderson's theorem. As we increase the scattering rate due to strong isotropic scatterers,  a low-energy impurity band is created in the density of states, show in Fig. \ref{DOSSpm}. The clean Raman response simply reflects two clean s-wave gaps with a sharp gap edge at $2\Delta_0$. Note the line-shape in a fully gapped superconductor. This qualitative feature seems differs from a nodal superconductors's line-shape, which still possesses a peak, only
more symmetric about the center energy at $2\Delta_0$.  One can understand the features in the Raman response as crudely similar to a convolution of the density of states with itself, so the effect of the impurity band on the Raman response is to create a nonzero threshold at $\omega<2\Delta_0$, as shown in Fig. \ref{RamanSpm}, corresponding to transitions between the impurity band and the gap edge.   There is, in addition, a very small contribution from scattering within the impurity band itself at low energies (see insert Fig. \ref{RamanSpm}).

\begin{figure}[h]
\includegraphics[width=0.75\columnwidth,angle=270]{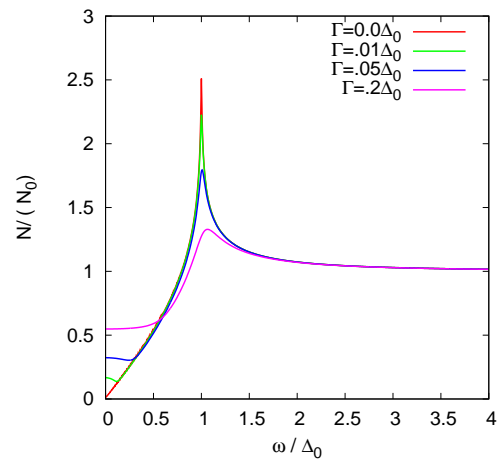}
\caption{(Color online) Density of states $N(\omega)/\Delta_0$ vs. energy $\omega/\Delta_0$ for a  $d$-wave superconductor for various values of scattering rate $\Gamma/\Delta_0$ in unitarity limit.  }
\label{DOSdwave}
\end{figure}

\begin{figure}[h]
\includegraphics[width=0.75\columnwidth,angle=270]{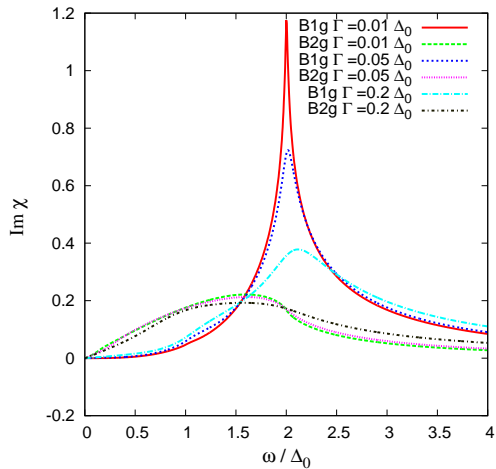}
\caption{ (Color online) Effect of disorder on $T=0$ Raman response of $d$ wave state vs. energy $\omega/\Delta_0$.  Shown are two polarizations, $B_{1g}$
and $B_{2g}$, for various values of scattering rate $\Gamma/\Delta_0$ in unitarity limit.}
\label{dwaveRaman}
\end{figure}

In Figs. \ref{DOSdwave} and \ref{dwaveRaman}, we show the corresponding quantities calculated in the simple 1-band $d$-wave case, $\Delta(\theta)=\Delta_0\cos(2\theta)$,
as a canonical example of what one expects for a nodal unconventional superconductor.
One fundamental difference with the $s_\pm$ case is the  presence of low energy quasiparticles in the clean limit. The nodes in the superconducting gap allow for  excitations  in the low-energy density of states, shown in Fig. \ref{DOSdwave}. Furthermore, the Raman polarization which weights  the nodes ($B_{2g}$, in the $d$-wave case) yields differing power laws in $\omega$ from the polarization which vanishes along the  nodal directions ($B_{1g}$), in contrast to the isotropic case where both of the Raman responses ($B_{1g}$ and $B_{2g}$) are the same. In the $d$-wave case, this is signaled by the presence ($B_{1g}$) or absence ($B_{2g}$) of a large peak at $2\Delta_0$. In both cases, increasing the scattering rate due to disorder increases the size of the impurity band, but the qualitative features of the $d$-wave Raman response, shown in Fig.  \ref{dwaveRaman} (a peak at $2\Delta_0$, excitations down to  $\omega=0$, differing response for differing polarizations) are unaltered by disorder. Beyond blurring of the sharp features, the effect of disorder will change the low frequency behavior between from $\omega^3$ to $\omega$ in the $B_{1g}$ polarization for the $d$-wave case\cite{DevKampf}.

\section{Raman response in models of ferropnictides}

\begin{figure}[h]
\includegraphics[width=0.75\columnwidth]{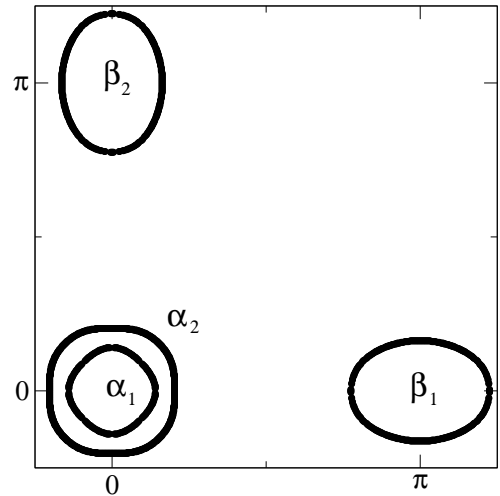}
\caption{ Fermi surface from a five-orbital tight binding model with parameters chosen to match the density functional theory result\cite{ref:Cao}.}
\label{FS}
\end{figure}

The full band structure of the Fe-pnictide LaOFeAs, determined by density functional calculations \cite{ref:Singh,ref:Cao}, can be accurately parameterized by a tight binding model with 5-bands \cite{s_graser_09}.
Generally, these materials have four Fermi surface sheets, shown in Fig. \ref{FS} in the ``unfolded" or 1-Fe zone: two hole pockets around the gamma point and two electron pockets. The two Fermi surfaces about the gamma point are referred to as $\alpha$ sheets and the other two are $\beta$ sheets. The exact details of the band structure are sensitive to doping \cite{kemper10}, and an additional hole pocket around the ($\pi,\pi$) point can occur. We do not include the ($\pi,\pi$)-pocket because its effect is expected to stabilize isotropic gaps in the context of multiorbital spin-fluctuation calculations\cite{k_kuroki_09,r_thomale_09,kemper10}, and because it occurs (within a rigid band shift implementation of doping) for the hole-doped cases only. We focus here on electron doped materials on which experiments have been performed, which appear to show nodes or deep gap minima.

To model the $A_{1g}$, $B_{1g}$, and $B_{2g}$ polarizations measured in Muschler et al. \cite{Muschler} the same choices were made for the extended s-wave gap on the $\alpha$ sheets (around the $\Gamma$ point) and $\beta$ sheets (around the M point) as in that paper:

\begin{eqnarray}
\Delta_{\alpha_1}(\theta) =\Delta_0 \frac{1+r \cos(4 \theta)}{1+r} & r= .75  \label{a1} \\
\Delta_{\alpha_2}(\theta)=\Delta_0 \frac{1-r \cos(4 \theta)}{1+r} & r= .75 \label{a2} \\
\Delta_{\beta_1}(\theta) =-\Delta_0 \frac{1-r \cos(2 \theta)}{1+r} & r= 1 \label{b1} \\
\Delta_{\beta_2}(\theta) = -\Delta_0 \frac{1+r \cos(2 \theta)}{1+r} & r= 1 \label{b2}
\end{eqnarray}

The vertices for this state are chosen to be:
\begin{eqnarray}
\alpha_1 &  \gamma_{1g} =0.&  \gamma_{2g} =0.  \\
\alpha_2 & \gamma_{1g}=.25( -2) \sin (\theta )\cos( \theta) &  \gamma_{2g} =.25 \cos(2 \theta) \\
\beta_1 & \gamma_{1g}= .5 ( -2) \sin (\theta )\cos( \theta)&  \gamma_{2g}=+1.\\
\beta_2 & \gamma_{1g}= .5 ( -2) \sin (\theta )\cos( \theta)&  \gamma_{2g} =-1.
\end{eqnarray}
Note that the notation is now in the 2-Fe zone, for easy comparison with Ref. \onlinecite{Muschler}.
The vertices have been chosen to model experimental results with several constraints in mind. First, we must respect the underlying symmetry of the polarization state. In the $B_{1g}$ polarization, the response weighted is largely away from any Fermi sheet, which is reflected by a flat response in the model and data. The $B_{2g}$ polarization
samples the electron ($\beta$) sheets\cite{Muschler}.  In the data and the model,
 there is a strong $T$-dependence to the data and a peak which appears below $T_c$, reflecting the corresponding gap function on these Fermi surfaces.   More realistic calculations of the
 vertex functions $\gamma_{\bf k}$  will be necessary for quantitative comparison with experiment.

\begin{figure}[h!]
\includegraphics[width=0.75\columnwidth,angle=270]{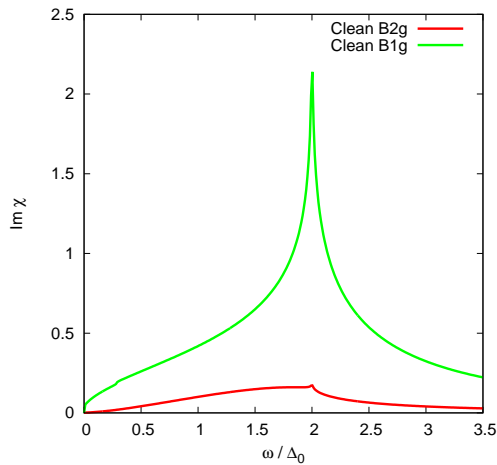}
\caption{ (Color online) Clean $B_{1g}$ and $B_{2g}$ spectra for the model of the superconducting state  defined in Eq. \ref{a1}-\ref{b2}\cite{Muschler} and in Eq. (\ref{a1}(-(\ref{b2}). }
\label{cleanXSram}
\end{figure}

 Figure \ref{cleanXSram} shows the clean results for this model. In Muschler et al.\cite{Muschler}, single crystals of Co doped \BFCA with x = 0.061 and 0.085 were studied. First, at x=0.061, the $B_{2g}$ spectra in the superconducting state are strikingly different from the $B_{1g}$ and $A_{1g}$ spectra, which is not possible for an isotropic gap. There is a sharp peak in the $B_{2g}$ polarization whose shape is not asymmetric as would be characteristic of a full gap, already suggesting the presence of nodes. The $B_{1g}$ polarization shows almost no change upon entering the superconducting state, which is consistent with vertices which probe the regions of the Brillouin zone without a Fermi surface. For \BFCA with x=0.061, there is a nonzero Raman intensity down to zero frequency, which indicates the presence of low energy quasiparticles. Thus far, this is reminiscent of the one-band d-wave model. To explore this further the low frequency behavior of the Raman intensity can be examined. It is straightforward to show that the observed ${\omega^{1/2}}$-dependence of the $B_{2g}$ is characteristic of a marginal or ``kissing" node, where the node just touches the Fermi surface. Examination at other dopings can now shed light on the nature of the state. A revealing aspect of the data\cite{Muschler} is the x=0.085 $B_{2g}$ spectra in the superconducting state. For this higher doping a finite gap of 10 cm$^{-1}$ can be resolved. The changing structure observed at two dopings suggests that impurities can play an important role.

\begin{figure}[h]
\includegraphics[width=0.75\columnwidth,angle=270]{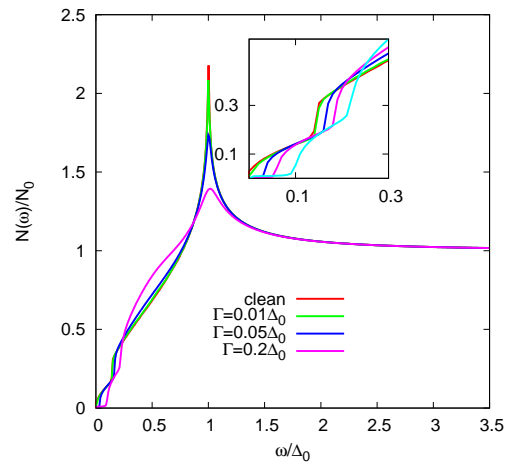}
\caption{(Color online) Density of states $N(\omega)/N_0$ for the extended s-wave state given in Eqs. Eq. \ref{a1}-\ref{b2} vs. $\omega/\Delta_0$  for unitary intraband scatterering rates $\Gamma/\Delta_0$.  Insert shows low-energy behavior.  }
\label{XSIntraDOS}
\end{figure}

\begin{figure}[h]
\includegraphics[width=0.75\columnwidth,angle=270]{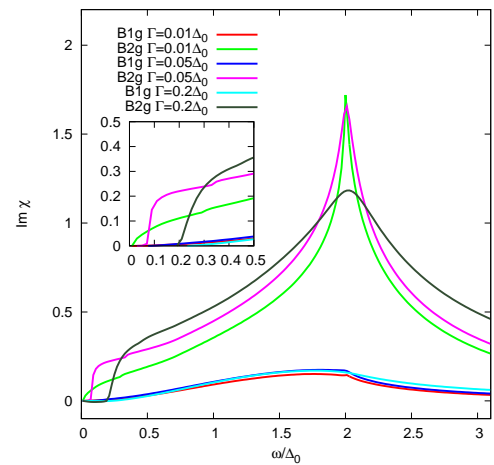}
\caption{(Color online)  Raman intensity for the extended s-wave state given in Eqs. Eq. \ref{a1}-\ref{b2} vs. $\omega/\Delta_0$  for various unitary intraband scatterering rates  $\Gamma/\Delta_0$ and polarization
states $B_{1g}$ and $B_{2g}$ in the 2-Fe zone.  Insert: low energy region. }
\label{XSintraRaman}
\end{figure}

The effect of disorder was calculated using the SCTMA in two limits: isotropic interband scattering for which $V({\bf k},{\bf k'})=V_0$, and intraband-only scattering for which $V({\bf k},{\bf k'})= 0$ if ${\bf k},{\bf k'}$ are on different Fermi sheets and $V_0$ if ${\bf k},{\bf k'}$ are on the same Fermi sheet.  It was assumed that the potential $V_0$ is large (unitarity limit), simulating the large density functional theory effective potential found for Co dopants in Ba-122\cite{AFKemperCo}.
In Fig.  \ref{XSIntraDOS} we can see the effect of increasing the intraband scattering rate on the density of states.
The gap $\Delta_{\beta_1,2}(\theta)$ has been chosen so that in the clean limit it has a  marginal or ``kissing" nodes leading to a characteristic $\omega^{1/2}$ behavior in the Raman response\cite{BoydRaman,Muschler}, and the asymptotic low-$\omega$ behavior of $N(\omega)$ is
indeed also $\omega^{1/2}$, as seen in Fig.  \ref{XSIntraDOS}; however there are additional features at low energies due to the
small minimum gaps on the hole sheets.  These features are largely suppressed in the clean Raman response, as shown in Fig. \ref{cleanXSram}.
In the presence of intraband disorder, the nodes on the $\beta$ sheets are lifted immediately, as seen in the density of states, creating a small but complete spectral gap $\Delta_{min}$, as discussed in Mishra et al.\cite{ref:vivek}  This, in turn, is reflected with the creation of a gap in the Raman intensity in Fig.  \ref{XSintraRaman}, where there is a disorder induced gap edge in the response up to $\omega=2\Delta_{min}$.

\begin{figure}[h]
\includegraphics[width=0.75\columnwidth,angle=270]{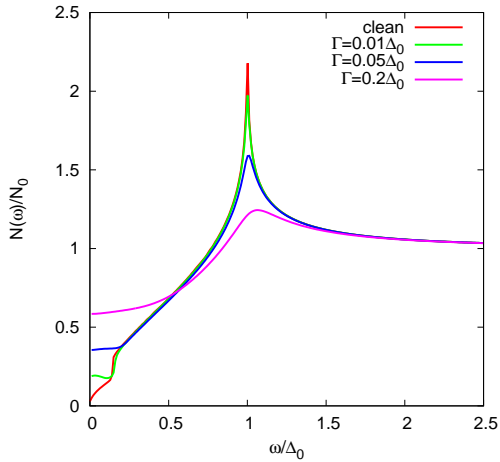}
\caption{(Color online) The density of states for an extended s-wave state given in Eqs. Eq. \ref{a1}-\ref{b2} vs. $\omega/\Delta_0$  for unitary isotropic scatterering rates $\Gamma/\Delta_0$. }
\label{XSInterDOS}
\end{figure}

\begin{figure}[h]
\includegraphics[width=0.75\columnwidth,angle=270]{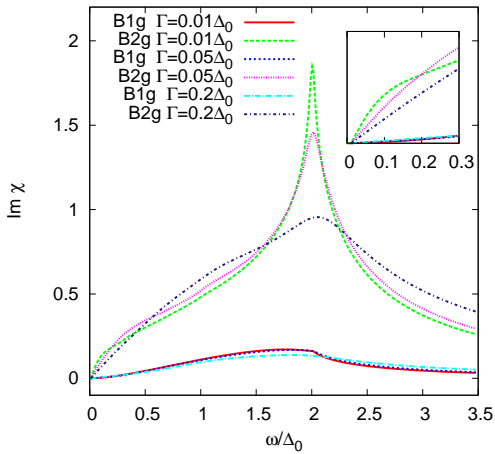}
\caption{(Color online) Raman intensity for an extended s-wave state of Fig. \ref{XSInterDOS} for various unitary isotropic scatterering rates  $\Gamma/\Delta_0$ and polarization
states $B_{1g}$ and $B_{2g}$ in the 2-Fe zone.  Insert: low-energy region.}
\label{XSinterRaman}
\end{figure}

The isotropic scatterers differ from the intra-band scatterers fundamentally in that inter-Fermi-sheet transitions allow for significant pair breaking, since the extended $s$-wave state changes average sign between electron and hole Fermi surfaces. So in addition to the gap-averaging that occurs as we scatter from ${\bf k}$ to ${\bf k'}$, there is the additional effect of creating a low energy quasiparticles ``impurity band." This effect prevents the lifting of nodes, so while the gap becomes more isotropic, the impurity band is the most important low energy effect in the density of states. Indeed, as shown in Fig.  \ref{XSInterDOS}, the DOS is reminiscent of the dirty $d$-wave case.   There is also a striking similarity in the Raman intensity shown in Fig. \ref{XSinterRaman} and the $d$-wave case. There are low energy quasiparticles all the way down to zero frequency, and we observe an enhancement of low frequency spectral weight.

\section{Conclusions}

Measurement of the Raman spectra in \BFCA revealed that at roughly 8\% doping there is a gap in the response up to  approximately 10 cm$^{-1}$, in contrast to the 6\% doped sample where there is a nonzero response down to zero frequency. This intriguing feature suggests a relation between cobalt doping, disorder, and the structure of the underlying superconducting gap. We have presented calculations of the Raman response as a function of frequency for varying
scattering rates due to disorder in the self-consistent T matrix approximation. First, we demonstrated the differing effects of
disorder on $s_\pm$ and $d$-wave gaps to illustrate how a nodal superconductor contrasts with a
sign-changing gapped state. The absence of low energy quasiparticles is an indication of a fully
gapped state, but for systems with interband disorder scattering  an impurity band is created which
can mimic nodal behavior. We found, however, that in the presence of disorder the form of the main
2$\Delta$ peak is largely unaffected; thus its observed symmetric form in the Muschler et al
experiment is suggestive evidence for true nodal or near-nodal behavior.

More definitive information is provided by the Raman response at low energies.  To study the
anisotropic $s$-wave state in more detail, we have presented a set of model gap functions on the 2D
Fermi surface sheets which capture the essential features of the experimental Raman response
\cite{Muschler}. It is immediately clear that for the dopings where measurements have been performed
a nodal state is likely because of two factors: low frequency power law behavior of the Raman
intensity down to zero energy, together with the very different responses for different
polarizations. We then examined different kinds of disorder on such a state. A remarkable effect is
the lifting of nodes by intraband scatterers for a highly anisotropic $s$-wave state, in rough
agreement with the qualitative effect of gapping the Raman spectrum observed as the doping increased
from 0.061 to 0.085 in Muschler et al. in \BFCA.  The predicted effects of strong interband
scattering at low energies were not observed, implying that the Co in this system acts
primarily as an intraband scatterer\cite{AFKemperCo}.  Of course, the present calculations include
only the effect of the impurity on pairbreaking, and neglect any direct effect on the pairing
interaction, so direct conclusions may not be drawn.

This feature distinguishes nodal anisotropic $s$-wave states from isotropic $s_\pm$ states in terms of how the response, and
underlying superconducting gap, evolves with increasing disorder. It also is not possible in
non-A$_{1g}$ representations of the gap, like d-wave, because the nodes cannot be removed for lower
symmetry superconducting gaps.
We believe that these distinctions will provide a useful basis for the interpretation of Raman
scattering experiments as doping or irradiation is tuned systematically in different samples.

 \acknowledgements  Research was partially supported by by DOE DE-FG02-05ER46236 (PJH) and  DE-AC02-76SF00515 (TPD).
The authors would like to acknowledge productive conversations with Dr.A.F.Kemper, Dr. R.Hackl, B. Muschler, and D.J. Scalapino.





\begin{thebibliography}{6}
\bibliographystyle{srt}


\expandafter\ifx\csname
natexlab\endcsname\relax\def\natexlab#1{#1}\fi
\expandafter\ifx\csname bibnamefont\endcsname\relax
  \def\bibnamefont#1{#1}\fi
\expandafter\ifx\csname bibfnamefont\endcsname\relax
  \def\bibfnamefont#1{#1}\fi
\expandafter\ifx\csname citenamefont\endcsname\relax
  \def\citenamefont#1{#1}\fi
\expandafter\ifx\csname url\endcsname\relax
  \def\url#1{\texttt{#1}}\fi
\expandafter\ifx\csname
urlprefix\endcsname\relax\def\urlprefix{URL }\fi
\providecommand{\bibinfo}[2]{#2}
\providecommand{\eprint}[2][]{\url{#2}}

\bibitem[{\citenamefont{DevereauxRMP}(2002)}]{DevereauxRMP}
\bibinfo{author}{\bibfnamefont{T.P.}~\bibnamefont{Devereaux}},
\bibinfo{author}{\bibfnamefont{R.}~\bibnamefont{Hackl}},
  \bibinfo{journal}{Rev.Mod.Phys.} \textbf{\bibinfo{volume}{79}},
  \bibinfo{pages}{175 } (\bibinfo{year}{2007}).


\bibitem[{\citenamefont{Muschler}(2009)}]{Muschler}
\bibinfo{author}{\bibfnamefont{B.}~\bibnamefont{Muschler}},
\bibinfo{author}{\bibfnamefont{W.}~\bibnamefont{Prestel}},
\bibinfo{author}{\bibfnamefont{R.}~\bibnamefont{Hackl}}
\bibinfo{author}{\bibfnamefont{T.P.}~\bibnamefont{Devereaux}}
\bibinfo{author}{\bibfnamefont{J.G.}~\bibnamefont{Analytis}},
\bibinfo{author}{\bibfnamefont{Jiun-Haw}~\bibnamefont{Chu}},
\bibinfo{author}{\bibfnamefont{I.R.}~\bibnamefont{Fisher}},
 \bibinfo{journal}{Phys.Rev.B} \textbf{\bibinfo{volume}{80}},
  \bibinfo{pages}{180510 } (\bibinfo{year}{2009}).

  \bibitem{DCJohnston} David C. Johnston arXiv:1005.4392 submited to advances in physics
  
\bibitem{PaglioneGreene} Johnpierre Paglione, Richard L. Greene arXiv:1006.4618 

\bibitem{kemper10} A. F. Kemper, T. A. Maier, S. Graser, H. Cheng,
P. J. Hirschfeld, and D. J. Scalapino, to appear in New J. of Phys.


\bibitem{ref:RKlingeler} R. Klingeler, N.~Leps, I.~Hellmann, A.~Popa, C.~Hess,
             A.~Kondrat, J.~Hamann-Borrero, G.~Behr, V.~Kataev, and B.~Buechner,
             arXiv:0808.0708.

\bibitem{ref:Grafe} H.-J.~Grafe, D.~Paar, G.~Lang, N.J.~Curro,
             G.~Behr, J.~Werner, J.~Hamann-Borrero, C.~Hess, N.~Leps,
             R.~Klingeler, and B.~Buchner, Phys. Rev. Lett. {\bf 101}, 047003 (2008).

\bibitem{ref:Ahilan} K.~Ahilan, F.L.~Ning, T.~Imai, A.S.~Sefat, R.~Jin, M.A.~McGuire,
B.C.~Sales, D.~Mandrus, Phys. Rev. B {\bf 78}, 100501(R) (2008).

\bibitem{ref:Nakai} T.Y. Nakai et al., J. Phys. Soc. Jpn. {\bf
77}, 073701 (2008).

\bibitem{ref:Zhao}L. Zhao et al Chin. Phys. Lett. {\bf
             25}, 4402 (2008).

\bibitem{ref:Ding} H.~Ding, P.~Richard, K.~Nakayama, T.~Sugawara, T.~Arakane, Y.~Sekiba, A.~Takayama,
             S.~Souma, T.~Sato, T.~Takahashi, Z.~Wang, X.~Dai, Z.~Fang, G.F.~Chen, J.L.~Luo, N.L.~Wang,
             Europhys. Lett. {\bf 83}, 47001 (2008).

\bibitem{ref:Kondo} T.~Kondo, A.F.~Santander-Syro, O.~Copie, C.~Liu, M.E.~Tillman, E.D.~Mun,
             J.~Schmalian, S.L.~Bud'ko, M.A.~Tanatar, P.C.~Canfield, A.~Kaminski,
             Phys. Rev. Lett. {\bf 101}, 147003 (2008).

\bibitem{ref:Evtushinsky} D.V.~Evtushinsky, D.S.~Inosov, V.B.~Zabolotnyy, A.~Koitzsch, M.~Knupfer,
             B.~Buchner, G.L.~Sun, V.~Hinkov, A.V.~Boris, C.T.~Lin, B.~Keimer, A.~Varykhalov, A.A.~Kordyuk, S.V.~Borisenko,
             arXiv:0809.4455.

\bibitem{ref:Nakayama}K. Nakayama, T. Sato, P. Richard, Y.-M. Xu, Y. Sekiba, S. Souma,
G. F. Chen, J. L. Luo, N. L. Wang, H. Ding, T. Takahashi,
arXiv:0812.0663.

\bibitem{ref:Hasan} L. Wray, D. Qian, D. Hsieh, Y. Xia, L. Li, J.G. Checkelsky, A. Pasupathy, K.K. Gomes,
C.V. Parker, A.V. Fedorov, G.F. Chen, J.L. Luo, A. Yazdani, N.P.
Ong, N.L. Wang, M.Z. Hasan, arXiv: 0812.2061.




\bibitem{ref:Hashimoto} K. Hashimoto, T. Shibauchi, T. Kato, K. Ikada,
 R. Okazaki, H. Shishido, M. Ishikado, H. Kito, A. Iyo, H. Eisaki, S. Shamoto, and Y. Matsuda arXiv:0806.3149.


 \bibitem{ref:Malone} L. Malone, J.D. Fletcher, A. Serafin, A. Carrington, N.D. Zhigadlo,
 Z. Bukowski, S. Katrych, and J. Karpinski,  arXiv 0807.0876.

 \bibitem{ref:Martin}C. Martin, R. T. Gordon, M. A. Tanatar, M. D. Vannette,
  M. E. Tillman, E. D. Mun, P. C. Canfield, V. G. Kogan, G. D. Samolyuk, J. Schmalian, and R. Prozorov,  ArXiv:0807.0876

 \bibitem{ref:Hashimoto2}K. Hashimoto etal. arXiv:0810.3506.

 \bibitem{ref:Gordon} R. T. Gordon, N. Ni, C. Martin, M. A. Tanatar, M. D. Vannette, H. Kim, G. Samolyuk, J. Schmalian,
  S. Nandi, A. Kreyssig, A. I. Goldman, J. Q. Yan, S. L. Bud'ko, P. C. Canfield, R.
  Prozorov, arXiv:0810.2295.

  \bibitem{ref:Gordon2} R. T. Gordon, C. Martin, H. Kim, N. Ni, M. A. Tanatar, J. Schmalian,
  I. I. Mazin, S. L. Bud'ko, P. C. Canfield, R. Prozorov,
  arXiv:0812.3683.

 \bibitem{ref:Fletcher} J.D. Fletcher, A. Serafin, L. Malone, J. Analytis, J-H Chu, A.S. Erickson, I.R. Fisher, A.
 Carrington, arXiv:0812.3858.


\bibitem{ref:Mazin_Spm}
 arXiv:0803.2740
    Title: Unconventional sign-reversing superconductivity in LaFeAsO1-xFx
    Authors: I.I. Mazin, D.J. Singh, M.D. Johannes, M.H. Du
        Journal-ref: Phys. Rev. Lett. 101, 057003 (2008)


\bibitem{s_graser_09} S. Graser, T. A. Maier, P. J. Hirschfeld, and D. J.
Scalapino, New. J. Phys. 11 (2009)

\bibitem{chubukov} A. V. Chubukov, D. V. Efremov, and I. Eremin, Phys. Rev.
B 78, 134512 (2008).

\bibitem{k_kuroki_08}
 K. Kuroki, S. Onari, R. Arita, H. Usui, Y. Tanaka, H. Kon-
tani, and H. Aoki, Phys. Rev. Lett. 101, 087004 (2008).

\bibitem{k_kuroki_09} K.~Kuroki, H.~Usui, S.~Onari, R.~Arita, and H.~Aoki,
             Phys. Rev. B {\bf 79}, 224511 (2009).

\bibitem{ref:Wang_nodal_gapped} F. Wang, H. Zhai, Y. Ran, A. Vishwanath, and D. Lee,
Physical Review Letters 102, 047005 (2009).

\bibitem{r_thomale_09}R. Thomale, C. Platt, J. Hu, C. Honerkamp, and B. A.
Bernevig, Phys. Rev. B 80, 180505 (2009).


 \bibitem{ref:vivek} V. Mishra, G.R. Boyd, S. Graser, T. Maier, P.J.
Hirschfeld, and D.J. Scalapino, Phys. Rev. B 79, 094512 (2009)

\bibitem{ref:Yu}
Yu, L., 1965, Acta Phys. Sin. 21, 75.
\bibitem{ref:Shiba}
Shiba, H., 1968, Prog. Theor. Phys. 40, 435.
\bibitem{ref:Hussey}
Hussey, N. Adv. Phys 51, 1685 (2002).

\bibitem{GolubovMazin}
A.A. Golubov and I.I. Mazin, Phys. Rev. B 55, 15146
(1997).

\bibitem{SengaKontani}
 Y. Senga and H. Kontani, arXiv:0809.0374;
arXiv:0812.2100. One of these is J. Phys. Soc. Jpn.
77, 113710 (2008).

 \bibitem{AFKemperCo} A. Kemper, C. Cao, P.J.
Hirschfeld, and H.-P. Cheng, Phys. Rev. B 80, 104511 (2009).

\bibitem{SawatzkyCo}H. Wadati, I. Elfimov, G. A. Sawatzky,
 arXiv:1003.2663.

\bibitem{Kempertobepublished} A.F. Kemper, M. Korshunov, and P.J. Hirschfeld, to be published.


\bibitem{DevKampf}T.P. Devereaux and A.P. Kampf  Int. J. Mod. Phys. B11, 2093 (1997).


\bibitem{DevEinzel}T.P. Devereaux and D. Einzel, Phys. Rev. B 51, 16336–16357 (1995).


\bibitem{BoydRaman} G.R. Boyd, T.P.
Devereaux,   P.J. Hirschfeld,  V.Mishra,
 and D.J. Scalapino, Phys. Rev. B 79, 174521 (2009).

\bibitem{TomMultiband} T. P. Devereaux, A. Virosztek, A. Zawadowski Phys.Rev.B 54, 12523 (1996)


\bibitem{TomPRL74} T. P. Devereaux    Phys.Rev.Lett. 74, 4313 (1995)

\bibitem{ref:Singh} D.J.~Singh and M.-H.~Du,  Phys. Rev. Lett. {\bf 100}, 237003 (2008).

\bibitem{ref:Cao} C.~Cao, P.J.~Hirschfeld, H.-P.~Cheng, Phys. Rev. B {\bf 77}, 220506(R) (2008).




\end{thebibliography}
\end{document}